# Exploring Non-Steady-State Charge Transport Dynamics in Information Processing: Insights from Reservoir Computing

Zhe-Yang Li,[1, 2] Xi Yu[1, 2*]

1. Key Laboratory of Organic Integrated Circuit, Ministry of Education & Tianjin Key Laboratory of Molecular Optoelectronic Sciences, Department of Chemistry, School of Science, Tianjin University, Tianjin 300072, China

2. Collaborative Innovation Center of Chemical Science and Engineering (Tianjin), Tianjin 300072, China

**ABSTRACT:** Exploring nonlinear chemical dynamic systems for information processing has emerged as a frontier in chemical and computational research, seeking to replicate the brain's neuromorphic and dynamic functionalities. We have extensively explored the information processing capabilities of a nonlinear chemical dynamic system through theoretical modeling by integrating a non-steady-state proton-coupled charge transport system into reservoir computing (RC) architecture. Our system demonstrated remarkable success in tasks such as waveform recognition, voice identification and chaos system prediction. More importantly, through a quantitative study, we revealed the key role of the alignment between the signal processing frequency of the RC and the characteristic time of the dynamics of the nonlinear system, which dictates the efficiency of RC task execution, the reservoir states and the memory capacity in information processing. The system's information processing frequency range was further modulated by the characteristic time of the dynamic system, resulting in an implementation akin to a 'chemically-tuned band-pass filter' for selective frequency processing. Our study thus elucidates the fundamental requirements and dynamic underpinnings of the non-steady-state charge transport dynamic system for RC, laying a foundational groundwork for the application of dynamic molecular devices for *in-materia* computing.

## Introduction

The exploration of nonlinear process and transient dynamics within chemical systems has consistently been at the forefront of scientific inquiry, given its intricate nature and the challenges it presents.[1, 2] A particular focus has been on nonlinear charge transfer process[3, 4] and non-steady-state dynamic systems[5] with evolving temporal intermediate states and dynamics, which culminate in the emergence of molecular devices with memory effects and dynamic switching capabilities.[6, 7]

Beyond their fundamental physical attributes, there has been a renewed interest in harnessing the nonlinear dynamic systems for information processing in the context of the burgeoning advancements in artificial intelligence and brain-inspired computation.[8-11] Unlike conventional transistors based digital computation with von Neumann architectures, where modern artificial neural network algorithms run, the human nervous system, exemplifying a unique computational model, stores and processes signals simultaneously through connected neurons.[12, 13] Moreover, it leverages elegant and sophisticated spatio-temporal dynamics,[14] which plays a key role for its capability of processing analog signals for cognitive learning and continuous adaptation with high efficiency and low energy consumption.[15, 16] There is, therefore, an increasing interest in the direct utilization of dynamic system with nonlinear dynamics to emulate the information processing functionality of the brain.[17-19] In this context, dynamic molecular devices, exhibiting temporal switching and memristive characteristics, emerge as promising physical systems to emulate the behavior of synapse and to realize the *in-materia* neuromorphic computing.[20-24]

Reservoir Computing (RC) is a subset of recurrent neural network: which utilizes the dynamics of a fixed, nonlinear reservoir system to map input signals into higher-dimensional computational spaces,[25-27] making it apt for tasks such as waveform recognition and time series prediction with high precision and resource-efficiency.[28, 29] It can be formulated in physical dynamical systems, which has positioned physical reservoir as a leading paradigm for exploiting high-dimensional, nonlinear, dynamical substrates for *in-materia* computation.[30-33] A defining feature of RC is its ability to execute neural network tasks with minimal dynamic devices, sometimes even a singular one, using the output of the dynamical physical system at different time step as nodes of the reservoir.[34] The minimal device requirement and high attribute demands of RC align seamlessly with nano and molecular scale devices, offering a promising avenue for designing novel computational paradigms. Reservoir computing (RC) has been well demonstrated in a series of devices like electrochemical transistor,[35-37] random connected nanowire network,[22, 38] and et. al.[39-44] Particularly, molecular junction of peptide monolayer device by metal ion migration induced nonlinear charge transport systems,[45] highlighting the potential of molecular-scale dynamic devices for information processing. The applications of RC for *in-materia* computation raised fundamental challenges concerning the underlining principle for RC to be realized by non-steady-state dynamic charge transport and the correlation between characteristics of the nonlinear chemical dynamic system and the information processing performance by RC.

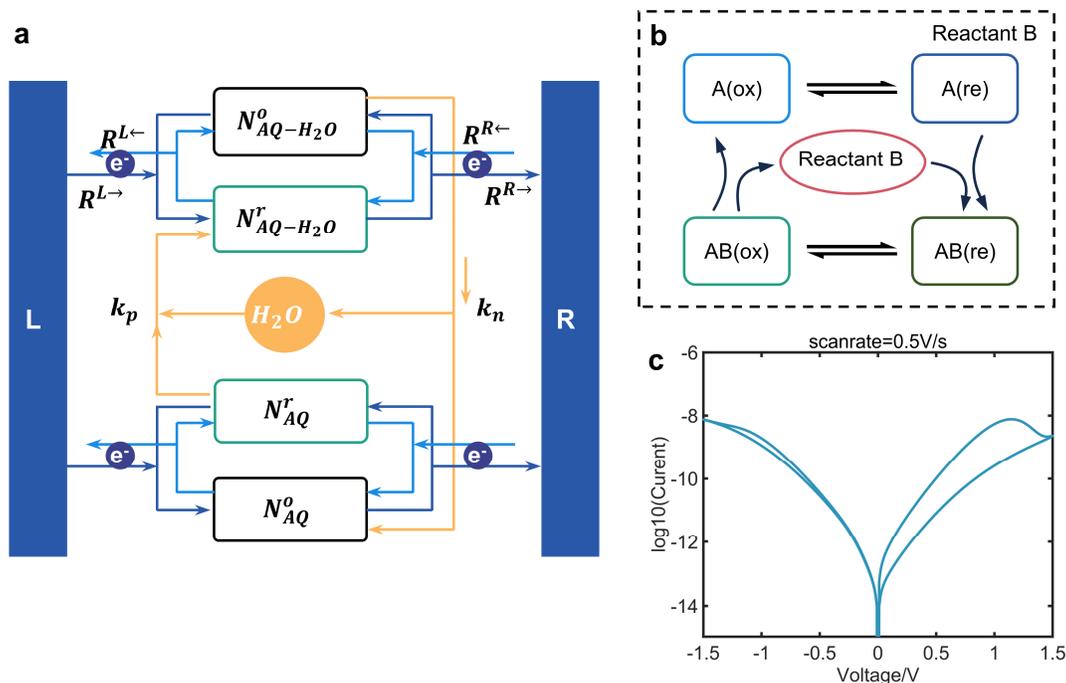

**Figure 1.** the mechanism and characteristics of the non-steady-state charge transport dynamics (a) The scheme of PCET dynamics (The water in the environment act as the proton source here). Electron transport through $AQ$ or $AQ - H_2O$ channel modulate the occupation of redox state, resulting in the proton transfer reaction and channels conversion (b) A general redox reaction coupling with electron transport dynamics. The reactants A and AB are redox active and environment reactant B is a part of reactant and holds a constant concentration. (c) A simulation current-voltage character in logarithm scale in a scan rate of $0.5V/s$.

In this study, we conducted an extensive theoretical modeling study of the information processing of RC, which was implanted in a nonlinear chemical dynamic system employing the proton-coupled non-steady-state charge transport dynamics, as its reservoir kernel. We found that the proton-coupled non-steady-state charge transport dynamic system is capable of supporting RC tasks effectively, achieving high performance waveform recognition,[46, 47] voice identification,[48] and nonlinear dynamic system behaviors prediction.[28] More importantly, we revealed that the alignment between the signal processing frequency of the RC and the characteristic time of the dynamic reservoir system dictates the efficiency of RC task execution, the reservoir states and the memory capacity in information processing. This research elucidates the fundamental requirements and dynamical underpinnings of the non-steady-state charge transport dynamic system for RC, laying a foundational groundwork for the application of dynamic molecular devices for in-*materia* computing.

## Dynamic Model and Reservoir Computing Algorithm

### The Proton-coupled non-steady-state electron transport system

The nonlinear electron transport (ET) system by proton-coupled non-steady-state charge transport has been well described in our previous study,[5] and we reiterate the fundamental idea here. As shown in figure 1a, the ET proceeds through a redox-active molecule anthraquinone ($AQ$), with intermediate state coupled to the proton transfer process, leading to temporally varying intermediate states and thus the non-steady-state transport dynamics.

The electron transfer cross molecule is often considered to be an instantaneous event, with traversal time on scale of ps, much faster than the perturbation voltage and measurement. Therefore, the electron (hole) transport can be regarded as a steady process, i.e.

$$N^o(R^{L\to} + R^{R\leftarrow}) - N^r(R^{L\leftarrow} + R^{R\to}) = 0 \quad (1)$$

Here $N^o$ and $N^r$ are the population of the oxidized and reduced state of the redox-active molecule, so that $N^o + N^r = 1$. $R^{L/R}$ is the charge transfer rate constant at the left and right molecule-electrode interface with arrow representing the direction of the charge flow. Therefore, the charge transport can be described by steady-state kinetic by assuming a time-invariant intermediate state population and the net current crosses the junction by ET is

$$I = -e(N^o R^{L\rightarrow} - N^r R^{L\leftarrow}) = -e(N^r R^{R\rightarrow} - N^o R^{R\leftarrow})$$
$$= -e\frac{R^{R\rightarrow}R^{L\rightarrow} - R^{L\leftarrow}R^{R\leftarrow}}{R^{L\rightarrow} + R^{R\leftarrow} + R^{L\leftarrow} + R^{R\rightarrow}} \quad (2)$$

The elementary charge transfer rate constant is described by Marcus theory as:

$$R^{L(R)} = \int_{-\infty}^{+\infty} dE\, k_{L(R)}(E)\, f_{L(R)}(E)\, F(\mu_{L(R)} - \Delta E) \quad (3)$$

In this context, $E$ represents the energy of the electronic states of the electrode. $k_{L(R)}(E)$ is the electron quantum transition rate between the left(right) electrode and the molecular state. $f(E)$ is the Fermi distribution of the electrode. Left or right electrode's electrochemical potential, $\mu$, is related to the bias voltage, $V_b$. Therefore, the charge transfer rate constants can be viewed as instantaneous response functions of $V_b$, thereby defining the current and the intermediate state population, which are represented as $R(V)$, $I(V)$, and $N(V)$.

The ET becomes non-steady-state as the intermediate state evolves with respect to time due to proton transfer, as shown in figure 1a. This process occurs on a timescale comparable to the measurements and variations in external stimuli. Through this proton transfer process, an additional ET channel appears via the redox of the protonated species. The total current of the junction is then the sum of the currents contributed by both channels.

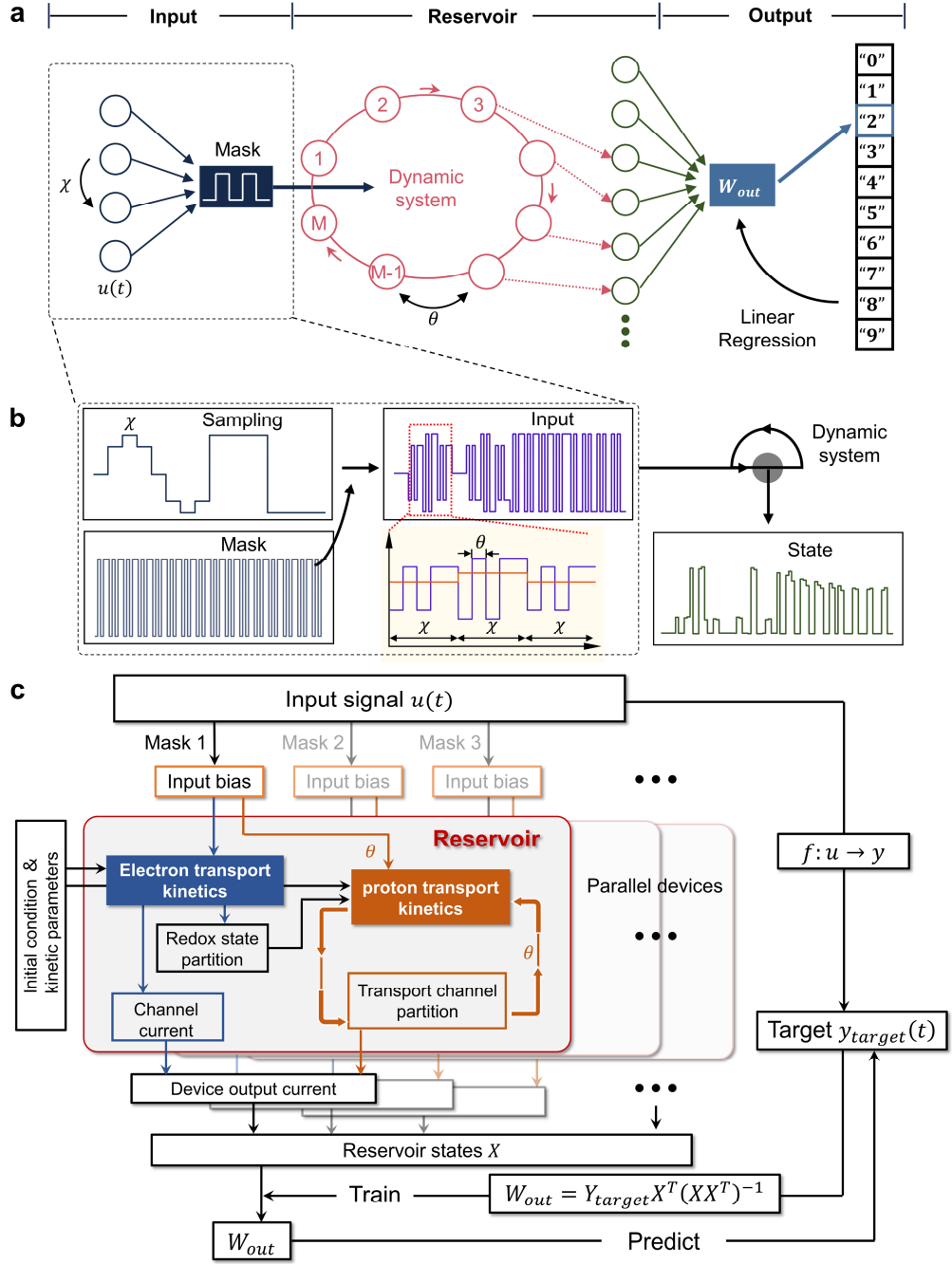

**Figure 2.** Schematic of the non-steady-state reservoir computing system. (a) The input signal $u(t)$ is multiplexed through a mask before entering the cyclic reservoir with $M$ virtual nodes. The output weight $W_{out}$ is trained using linear regression and functions as the weight matrix to map the reservoir states to the output. (b) The masking procedure is as follows: an input signal is sampled over a duration $\chi$ and then multiplexed through a random mask. The resulting temporal sequence is then fed into the virtual nodes at intervals of $\theta$, thus generating the reservoir states necessary for the training phase. The insert panel with yellow background shows a schematic of the masked input stream. The non-steady-state dynamic response during the duration $\chi$ comprises $M$ virtual nodes, each with a timestep of $\theta$.

$$I_{total} = Q_{AQ}I_{AQ} + Q_{AQ-H_2O}I_{AQ-H_2O} \quad (4)$$

Here $Q_{AQ/AQ-H_2O}$ represents the population of the $AQ$ and $AQ-H_2O$ channels, such that $Q_{AQ} + Q_{AQ-H_2O} = 1$ and $N^o_{AQ/AQ-H_2O} + N^r_{AQ/AQ-H_2O} = Q_{AQ/AQ-H_2O}$. Note here that the steady-state electron transport assumption only ensures the relative population $N^o_{AQ/AQ-H_2O}$ and $N^r_{AQ/AQ-H_2O}$, i.e. the ratios $N^o_{AQ}$ and $N^r_{AQ}$ are functions of bias voltage, similar to $AQ - H_2O$. However, these ratios are time variant due to the proton transfer kinetics and this slow process is the origin of the non-steady-state charge transport, as described by

$$I_{total}(t) = Q_{AQ}(t)I_{AQ} + Q_{AQ-H_2O}(t)I_{AQ-H_2O} \quad (5)$$

The kinetics of the reversible reaction between the two channels provides the time-dependent variation of $Q_{AQ/AQ-H_2O}$ by

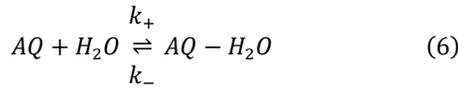

$$AQ + H_2O \underset{k_-}{\overset{k_+}{\rightleftharpoons}} AQ - H_2O \quad (6)$$

where $k_+$ and $k_-$ are the forward and backward reaction rate constant. The kinetics of this reaction is given by:

$$\frac{dQ_{AQ}}{dt} = -k_p Q_{AQ} + k_n Q_{AQ-H_2O} \quad (7)$$

Here $k_p(V, H_2O) = N^r_{AQ}(V)Q_{H_2O}k_+$ and $k_n(V) = N^o_{AQ-H_2O}(V)k_-$ are the effective rate constants that incorporate the concentration of $H_2O$ and the bias voltage dependent population of the reactive species, $N^r_{AQ}(V)$ and $N^o_{AQ-H_2O}(V)$ of each channel at the steady-state electron transport (see *Supporting Information* for details).

We then obtained the time dependent channel population:

$$Q_{AQ}(t) = \left(Q_{AQ}(0) - \left(\frac{k_n}{k_p + k_n}\right)\right)e^{-(k_p + k_n)t} + \frac{k_n}{k_p + k_n} \quad (8)$$

The total current of the device will be obtained by incorporating $Q_{AQ}(t)$ and $Q_{AQ-H_2O}(t)$ into Eq 5.

In the operation of the junction device, the applied bias voltage becomes temporal, so that the electrical response characteristics of this delayed system have to be obtained recursively. The system's output at any given time affects subsequent outputs with a delay time step $\Delta t$.

$$Q_{AQ}(t) = \left[Q_{AQ}(t - \Delta t) - \left(\frac{k_{n,t-\Delta t}}{k_{p,t-\Delta t} + k_{n,t-\Delta t}}\right)\right] \times$$
$$e^{-(k_{p,t-\Delta t} + k_{n,t-\Delta t})\Delta t} + \frac{k_{n,t-\Delta t}}{k_{p,t-\Delta t} + k_{n,t-\Delta t}} \quad (9)$$

$k_{p,t}$ and $k_{n,t}$ reflect the time dependency of the bias voltage. The current of the device, given in a recursive manner, is:

$$I(t) = I_{AQ-H_2O}(V(t)) + \left(I_{AQ}(V(t)) - I_{AQ-H_2O}(V(t))\right) \times$$
$$\left[\frac{I(t - \Delta t) - I_{AQ-H_2O}(V(t))}{I_A(V(t)) - I_{AQ-H_2O}(V(t))} - \left(\frac{k_{n,t-\Delta t}}{k_{p,t-\Delta t} + k_{n,t-\Delta t}}\right)\right] \times$$
$$e^{-(k_{p,t-\Delta t} + k_{n,t-\Delta t})\Delta t} \quad (10)$$

This delayed system results in a time and history dependent I-V response with hysteresis, negative differential resistance, and memory effect.

It is worth mentioning that the nonlinear non-steady-state dynamics of the proton-coupled electron transport can be generalized to any reversible chemical reaction, as shown in figure 1b, although we discuss only the proton-coupled electron transport here.

## Implementation of reservoir computing

Reservoir computing (RC) can be implemented within a nonlinear dynamic system, where the states of the system at different time intervals work as nodes of the reservoir[34]. The nonlinear property of the system maps the input signal into a higher dimensional space where the signal become linearly separable. During the learning process, only the output mapping from the reservoir to the target is trained, using linear regression,[49] which is considerably simpler than training traditional recurrent neural networks. Furthermore, the dynamic system's inherent delayed memory characteristic is crucial for processing temporal sequences where the signal's history is of importance.

The proton-coupled electron transport non-steady-state dynamics is integrated into the reservoir computing framework. Figure 2a shows the reservoir computing system based on single dynamic device, consisting of an input layer, a reservoir layer and an output layer. The input layer processes the input information through masking process.[50] The reservoir layer, central to the system, generates numerous reservoir states through the non-steady-state electron transport dynamics, enabling classification. Time-multiplexed sequential signals from the input layer establish virtual nodes within the reservoir layer. By leveraging the memory effect of the non-steady-state dynamical system, these virtual nodes become interconnected.

**Table 1. Parameters for non-steady-state dynamic system reservoir computing tasks.**

| | | Waveform recognition | Digit recognition | Time-series prediction | NARMA-10 prediction |
|---|---|---|---|---|---|
| Reservoir computing parameters | Mask length $M$ | 5 | 10 | 5 | 5 |
| | Number of parallel devices $N$ | 8 | 40 | 25 | 8 |
| | Time interval $\theta$/s | 0.2 | 0.2 | 0.05 / 2.5 | 0.2 |
| | Input frequency $f$/Hz | 1 | 0.5 | 4 / 0.08 | 1 |
| Proton-coupled non-steady-state charge transport parameters | $k_+$ | 10 | | | |
| | $k_-$ | 0.05 | | | |
| | Water concentration $Q_{AQ-H_2O}$ | 20% | | | |

We introduce the input signal by the bias voltage into our non-steady-state dynamic system. The voltage sequence, with a time interval of $\theta$, is set to be the same as the delay time step $\Delta t$ in the non-steady-state charge transport dynamic model. The time interval $\theta$ or input frequency $f$ will be used to describe the temporal character of RC.

$$f = \frac{1}{\chi} = \frac{1}{M \times \theta} \quad (11)$$

Here $M$ is the mask length.

Figure 2b exemplifies the reservoir computing process using harmonic and square waveform recognition task. The randomly generated waveform is first sampled at 8 signal points in each period. The sampling procedure is controlled in relation to the raw signal frequency, ensuring a fixed number of sampling points per period. Consequently, a higher raw signal frequency results in a shorter sampling interval time $\chi$, as depicted in the upper left panel of figure 2b. The mask is a sequence of random binary value with the length (mask length) $M$ times the total sampling points ($P$) as $M \times P$. The upper middle panel of figure 2b shows the time-multiplexed inputs by the mask, where the time intervals are $\theta = \chi/M$, as depicted in the insert panel. Consequently, sequential signals traverse virtual nodes, generating a diverse array of reservoir states. During the training phases, the weight matrix $W_{out}$ is determined by: $W_{out} = Y_{target} X^T (XX^T)^{-1}$ to map the reservoir states onto the target using linear regression, as illustrated in figure 2a. The superscript -1 denotes the Moore-Penrose pseudoinverse. The linear training and testing process largely simplify the neural network computation. The algorithmic schematic is showcased in figure 2c.

To ensure optimal operation of the non-steady-state dynamic reservoir, we normalized the input voltage signal within the range of $V_{min} = -0.5V$ to $V_{max} = 1.5V$.

$$V = \frac{S - S_{Min}}{S_{Max} - S_{Min}} \times (V_{Max} - V_{Min}) + V_{Min} \quad (12)$$

Here $S$ is the input sequence.

## Results and Discussion

### Waveform and digit recognition

Classification and recognition are fundamental functions achievable through reservoir computing. To validate the basic reservoir properties of the non-steady-state dynamic systems, we perform waveform and voice recognition tasks. Throughout the session, the parameters of the experiments are depicted in table 1.

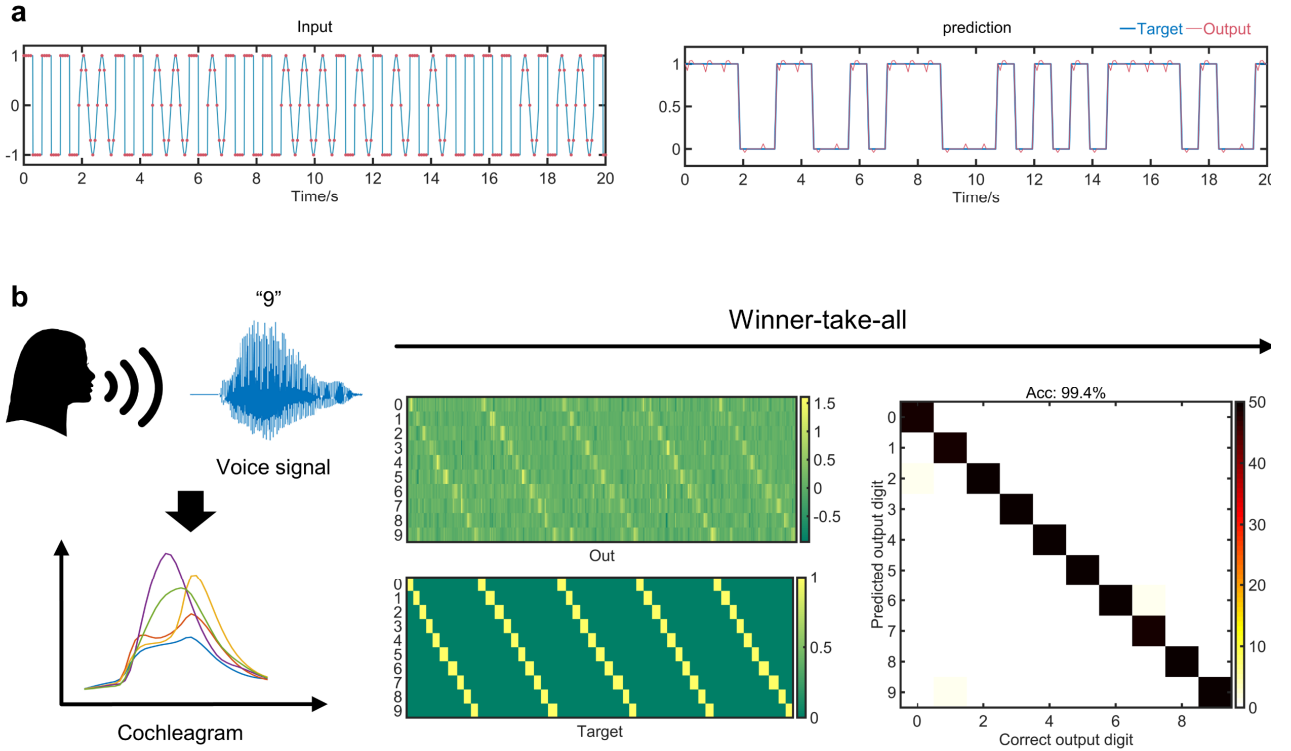

**Figure 3.** Classification task demonstrations by the non-steady-state dynamic RC. (a) Waveform recognition task: The graph presents both the input signal, a random mixture of harmonic and square waves, alongside the corresponding predictions. The target output is binary, with 0 representing harmonic waves and 1 representing square waves. The blue line indicates the raw signal, and red point represent the sampling data in the left panel. In the right panel, the blue and red lines represent the desired target and the actual output, respectively. (b) Spoken-digit recognition task: The diagram outlines the conversion of voice signals into a cochleagram across numerical frequency channels, which the reservoir then processes. The adjacent panel displays the output and target of the non-steady-state dynamic RC. A value of 1 indicates the trained digit within every vertical vector. The prediction's recognition accuracy rate reaches 99.4% using the winner-take-all strategy. The color bar reflects the total predicted

For the waveform recognition task, square waves were labeled as 1, and harmonic waves as 0. The results, depicted in figure 3a, show the original waveform signals and their sampling points in the left panel. The right panel demonstrates a clear classification of the two waveforms, yielding a normalized root mean square error (NRMSE) of ~0.01. We used a mask length of five and replicated the computations eight times within the reservoir to model the state matrix, equivalent to eight parallel dynamic devices.

For audio digit recognition, we processed NIST TI-46 spoken digit recordings with a Lyon passive ear model filter to generate cochleagram. The dataset included ten digits articulated by five speakers, divided into 450 training and 50 testing samples. Each digit was represented by a ten-dimensional binary vector. Classification was accomplished using 40 parallel devices, employing a winner-take-all strategy as detailed in figure 3b.

The overall prediction accuracy reaches an impressive 99.4%, confirmed through 10-fold cross-validation. This high accuracy underscores the effective information processing capabilities of the proton-coupled non-steady-state charge transport dynamics utilized. This success is attributed to the precise interplay between computational masks, input frequency, and dynamic constants integral to the system's efficiency. [28] Crucially, our results indicate that aligning the RC's processing frequency with the dynamic system's characteristic time is essential for optimal performance in waveform recognition and voice identification, highlighting the dynamic reservoir system's capacity for complex information processing tasks, which will be discussed later.

### Time-series prediction

Predicting chaotic sequences is a benchmark for assessing a reservoir's capability in time-series tasks. Chaotic systems, such as the Mackey-Glass system,[51] the Hénon map system[52] and the Lorenz weather system,[53] are commonly utilized for this purpose. These systems are sensitive to their history, and the objective is to predict the system's future state at time step n+1 based on its current state at time step n. The Hénon map, a typical discrete-time chaotic system, applies a nonlinear 2-D transformation to determine a point's new position $x(n+1), y(n+1)$ on a plane, defined as follows:

$$x(n+1) = y(n) - 1.4x(n)^2$$
$$y(n+1) = 0.3x(n) + w(n) \tag{13}$$

where $w(n)$ is the noise with 0 mean value and 0.05 standard deviation.

We compiled a dataset of 4000 points, splitting it evenly between training and testing, and employed a masking process similar to that used in the waveform recognition task. The system performed notably, with precise time-series predictions for the $x$ variable and successful attractor reconstructions as shown in figure 4a. Superior performance was observed at discrete time intervals $\theta$ of 2.5 s than that of 0.05 s.

The Lorenz system is characterized by the following differential equations:

$$\dot{x}_1 = \sigma(x_2 - x_1)$$
$$\dot{x}_2 = x_1(\rho - x_3) - x_2$$
$$\dot{x}_3 = x_1x_2 - \beta x_3 \tag{14}$$

We adopted the classical parameters $\sigma = 10, \rho = 28, \beta = 8/3$.[53]

A sequence of 2000 steps was computed from initial conditions, with the first half for training and the second for testing. Utilizing the same dynamic and reservoir parameters optimized for the Hénon map, the non-steady-state dynamic RC system demonstrated robust predictive capabilities, accurately reconstructing the intricate Lorenz attractor showcased in figure 4b.

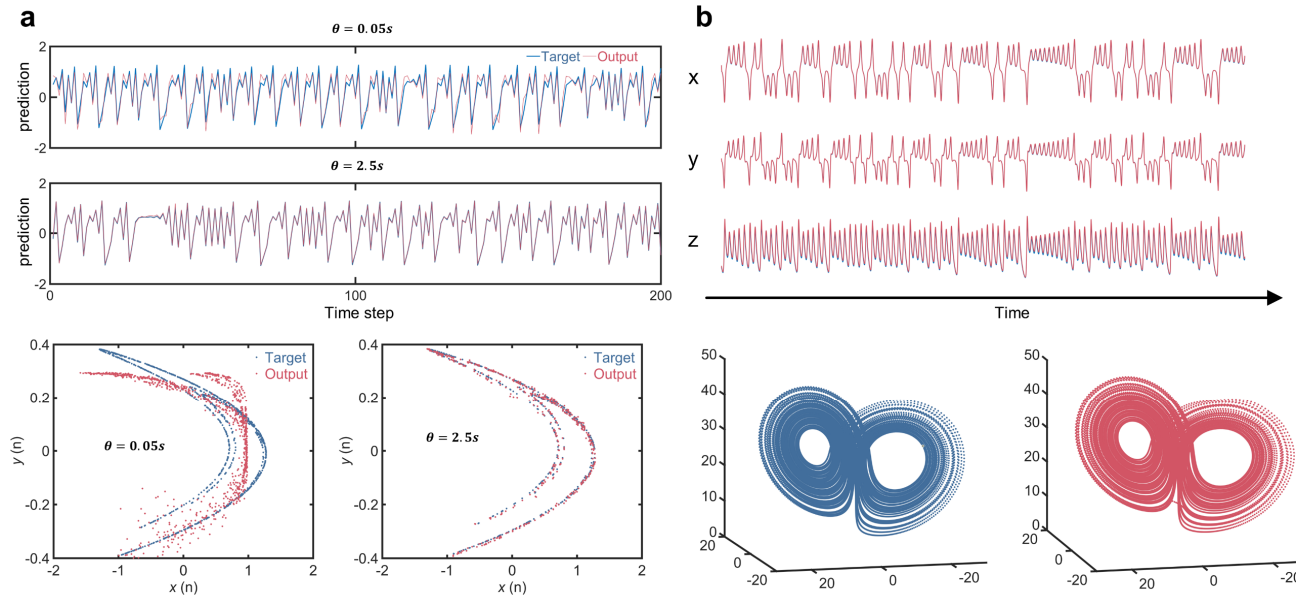

**Figure 4.** Prediction tasks demonstrations by the non-steady-state dynamic RC (a) Hénon map task: The predicted values of $x(n)$ and a two-dimensional plot of the intended targets (in blue) with the RC-generated outputs (in red). The results indicate that a temporal interval $\theta$ of 0.05 s yields suboptimal predictions when compared to an interval $\theta$ of 2.5 s. (b) Lorenz atmospheric task: The predicted results and a three-dimensional visualization that showcasing accuracy of the RC system in replicating the complex patterns of the Lorenz weather model.

This study affirms strong potential of the non-steady-state dynamic systems in reservoir computing, particularly in signal recognition and time-series prediction tasks. It highlights the importance of fine-tuning the RC system's parameters, like the number of parallel reservoirs $N$ and the mask length $M$, for optimal performance. While increasing the number of parallel reservoirs enhances the state dimension at the expense of computational complexity, a typical mask length of around five extends the state space in another dimension.[50]

The focus of subsequent research will be to discern how the kinetics of the non-steady-state dynamical system interact with RC parameters to achieve superior performance, a study that inherently bridges chemical kinetics and information processing.

## Quantitative analysis of RC performance dependent on the nonlinear kinetics

We conduct a rigorous evaluation of non-steady-state dynamic systems as reservoirs, focusing particularly on the role of input frequencies in reservoir computing. The introduced frequency is crucial as it governs the RC system's temporal memory and information processing capabilities. Achieving resonance between the reservoir's dynamics and the input frequency is critical for proficient signal processing and feature extraction, ensuring system stability and enabling the necessary nonlinear transformations.

To understand the RC system's performance with respect to the characteristics of non-steady-state dynamics, we examine the system's response to input signals over a range of frequencies $f$ $\left(f = \frac{1}{\chi} = \frac{1}{M \times \theta}\right)$. The recognition rate is chosen as the primary metric to measure the accuracy of the RC system in recognition experiments, indicative of its overall efficacy. Additionally, we assess three other parameters integral to the non-steady-state dynamics:

**feedback strength**, indicating the system's response intensity,

**state richness**, denoting the diversity of responses,

**memory capacity**, reflecting the system's ability to preserve and utilize previous information.

The waveform recognition task provides a platform for evaluating above parameters, offering definitive input frequencies for analysis. The 10th-order nonlinear autoregressive moving average (NARMA-10) task, a complex standard benchmark, was employed to gauge the reservoir's memory capacity. Our analysis seeks to unravel the influence of input frequency on these four parameters, and then to correlate RC computational results with the dynamic characteristics of the non-steady-state system, identifying the key factors that affect the core dynamics critical to RC functionality. Throughout these experiments, which span a broad frequency spectrum, we maintain parameters consistent with those established for waveform recognition task.

## Dynamic characteristics of the non-steady-state system

To assess RC performance, we conducted the waveform recognition task 44 times, analyzing recognition outcomes statistically across varying input frequencies, as illustrated in figure 5a. Identifications with a NRMSE below 0.1 were deemed successful. The recognition rate—normalized counts of successful recognition—shows robust performance from 0.1Hz to 10Hz, with a noticeable decline observed beyond this range.

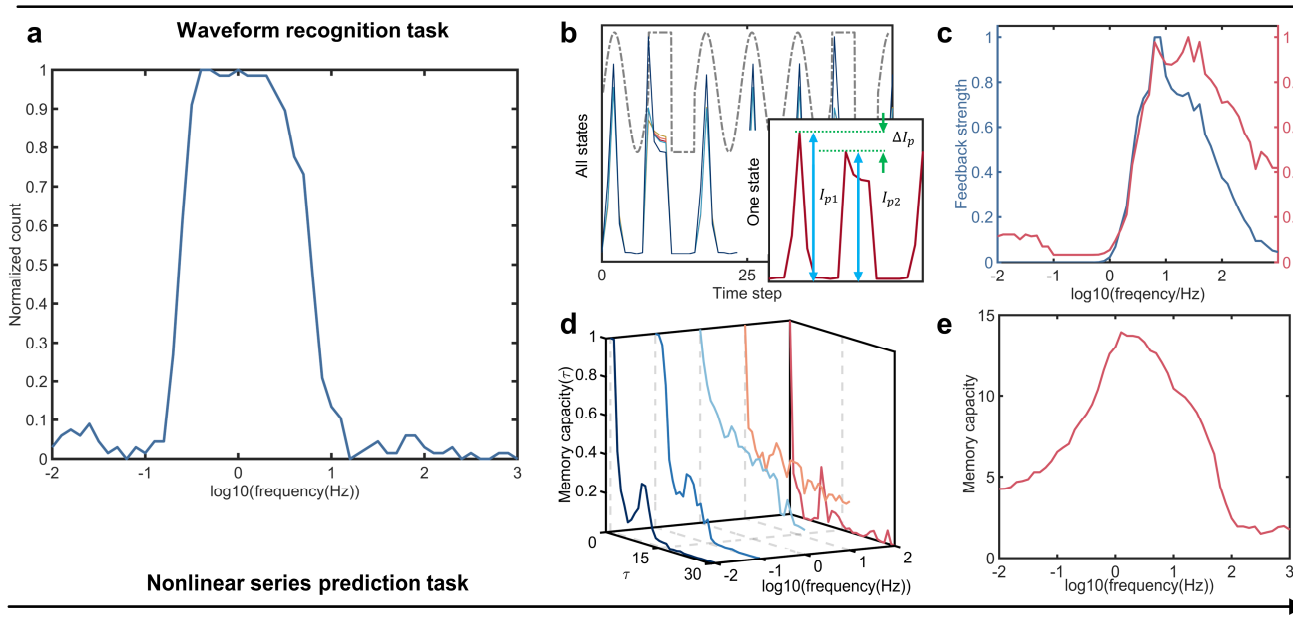

**Figure 5.** Quantitative analysis of RC performance and dynamic characteristics in the non-steady-state system. (a) Normalized successful recognition counts for 44 waveform recognition tasks on a logarithmic scale and peak at $f = 10^0\,\mathrm{Hz}$. (b) Five reservoir states observed during the waveform recognition task. The raw waveform is represented by a gray line, and an inset panel provides a detailed view of a specific state between time steps 30 to 50. (c) Feedback strength (in blue) and states richness (in red) plotted against logarithmic frequency, both reaching a peak around $f = 10^1\,\mathrm{Hz}$ (d) Memory capacity is evaluated under five distinct frequency conditions, with each condition's plot line demonstrating variation in relation to the delayed time step $\tau$. (e) The overall memory capacity, plotted against logarithmic frequency, achieves its peak around $f = 10^0\,\mathrm{Hz}$.

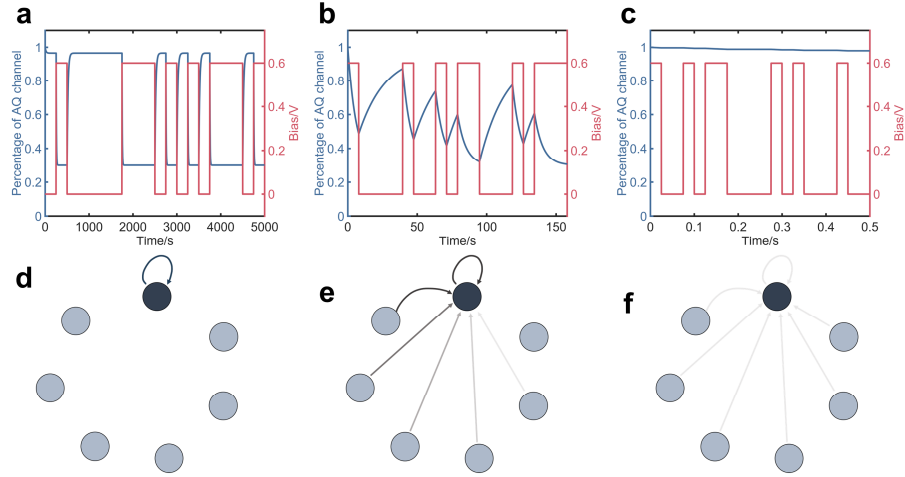

**Figure 6.** Schematic of temporal dynamic and interaction structure. Time evolution of AQ percentage (in blue) and stimulus voltage (in red line) over 500 timesteps (a, b, c), alongside the associated interaction structures at various input frequencies (d, e, f). (a) and (d) exhibit equilibrium dynamic at a low frequency ($f = 10^{-1}$Hz), where the virtual nodes show self-coupling. (b) and (e) display the non-steady-state dynamic at an intermediate frequency ($f = 10^{0.5}$Hz), with virtual nodes demonstrating coupling to their neighbors and to themselves. (c) and (f) reveal near-equilibrium dynamics at a high frequency ($f = 10^3$Hz), with virtual nodes exhibiting weak coupling throughout the network.

In waveform recognition, the dynamic system's current response signifies the reservoir state, where $I_{p1}$ and $I_{p2}$ denote peak responses to harmonic and square waves, respectively. The differences in stability and intensity of these responses to multiple stimuli offer insight into the reservoir's state richness and feedback capabilities.[28, 34] Time multiplexing of the input signal generates $M$ distinct reservoir state groups, as shown in figure 5b. Feedback strength $\overline{\delta_{I_p}}$ is quantified by averaging the five largest peak current differences ($\delta_{I_p}$) across all states, and state richness is measured by the standard deviation of the harmonic peak responses shown as: $\sigma_{I_{p1}} = \sqrt{\frac{1}{n}\sum_{i=1}^{n}(I_{p1,i} - \overline{I_{p1}})}$. As shown in figure 5c, feedback strength and state richness both peak around 10 Hz within a bandwidth approximately spanning an order of magnitude. This suggests that proper frequency of the input signal is important for the reservoir's richness and feedback strength.

Fading memory is a critical property of the dynamical system serving as reservoir, indicating the correlation between current and previous states at specific nodes. Memory capacity (MC) quantitatively captures the historical dependence of the reservoir's dynamics through a reconstruction of a nonlinear sequence task at a given delay step.[54, 55] In reservoir computing, the short-term MC with a delay step of $\tau$ is defined as:[56]

$$MC(\tau) = \frac{\text{cov}^2(x(t-\tau), y(t))}{\text{var}(x(t))\text{var}(y(t))} \quad (15)$$

where $x(t)$ is the input signal, and $x(t-\tau)$ is a delayed version of the nonlinear sequence with delayed time step of $\tau$, serving as the target signal, while $y(t)$ is the output of the reconstruction task. 'cov' and 'var' denote covariance and variance.

At $\tau = 0$, $MC(\tau)$ represents self-sequence prediction, typically yielding an $MC(0)$ value of 1, indicating perfect predictive correlation. For other time step delays $\tau$, the correlation coefficients range between 0 and 1, with MC diminishing towards 0 as $\tau$ increases. The overall MC is expressed as:

$$MC = \sum_{k=1}^{\infty} MC(\tau) \quad (16)$$

As proposed by Jaeger,[56] the reservoir is expected to predict a completely random sequence. While in practical situations, input sequences often comprise a mixture of structured signals and noise. To reflect this, we use the NARMA-10 system as a widely recognized benchmark for MC detection, to simulate situations where regular and noisy signals are intertwined,[57-59]

$$x(t+1) = 0.3x(t) + 0.05x(t)\sum_{i=0}^{9}x(t-i)$$
$$+ 1.5n(t)n(t-9) + 0.1 \quad (17)$$

where $n(t)$ is random white noise within [0, 0.5].

Figure 5d depicts MC of RC system while handling input signals across a spectrum of frequencies, showing a gradual decline to zero around a delay time step of $\tau = 15$. Figure 5e presents a global view of MC in relation to input frequency, with MC peaking at an input frequency of $10^0$Hz, signifying the most pronounced fading memory effect.

Comparing with recognition rate, both feedback strength and state richness demonstrate peaks at input frequencies nearly an order of magnitude higher, due to high-frequency signals inducing more distinct reservoir state variations. However, at ultra-high frequencies, differentiation of reservoir states diminishes. Global memory capacity metric emerges as the most consistent and effective indicator of the recognition performance for the non-steady-state

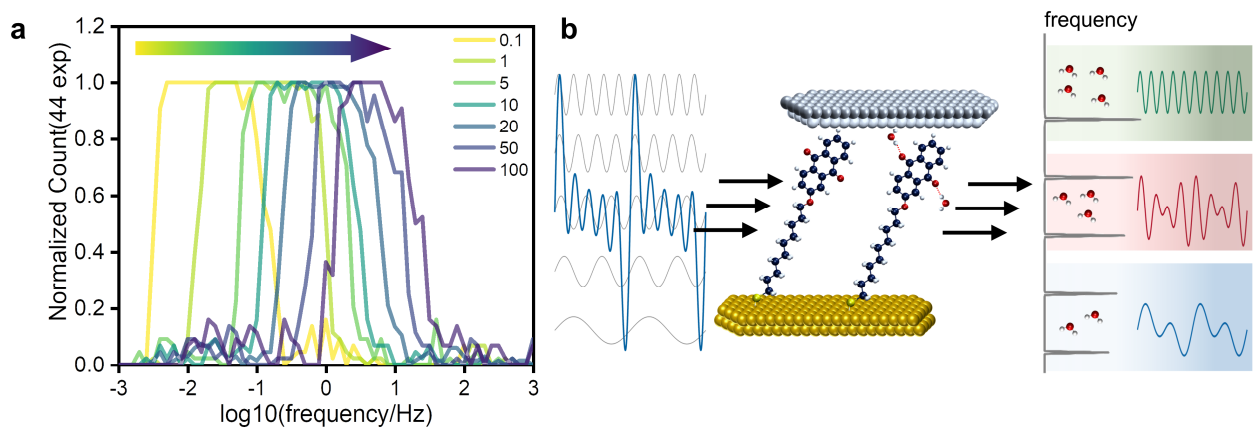

**Figure 7.** Humidity dependent experiments of the non-steady-state dynamic RC. (a) Normalized counts of successful waveform recognition across varying input frequencies. Environmental water concentration dynamically modulates peak shifts in recognition performance from 0.1 to 100% relative humidity, maintaining a consistent trend. (b) Schematic of the environmentally controlled non-steady-state dynamic device for neural network recognition tasks. The mixed wave is fed into the physical RC system, resulting in a filtering effect.

dynamic RC systems, peaking at the same input frequency.

In the context of the proton-coupled non-steady-state dynamic system, the primary cause of state changes in such a reservoir is the conversion between $AQ$ and $AQ - H_2O$. For an in-depth analysis of the system's dynamical properties, we examine the ET channels and the network of connections among nodes within the reservoir, focusing specifically on their responses to signals at three distinct frequencies.

Figure 6a presents the percentage of $AQ$ channels, which indicates the proportion of ET channels passing through $AQ$ in relation to the total channels, at a relatively low input frequency of $10^{-1}$Hz. At this frequency, far below the system's characteristic frequency, the uniformity across nodes leads to a lack of significant memory effects, posing a challenge for effective reservoir computing. The independence of the nodes is graphically depicted in figure 6d, with each node operating independently.

Figure 6b shifts focus to the system's behavior at a frequency of $10^{0.5}$Hz. At this frequency, we observe notable variations in species concentration at intervals, indicating the system does not reach equilibrium and aligns with favorable conditions for reservoir computing. Figure 6e further reveals that at this optimal frequency, each node not only self-connects but also links to a select number of preceding nodes, with the link strength weakening over distance.

The behavior under ultra-high-frequency input of $10^3$Hz is examined in figure 6c. In this scenario, the minimal changes in reservoir states at each step result in inadequate enhancement of the system's dimensionality for classification tasks, coupled with a reduced memory effect. Figure 6f then illustrates the virtual nodes when subjected to these ultra-high-frequency input. A network of weak interconnections among all nodes is observed, which emphasizes the system's limitations under such extreme conditions.

## Chemical modulation of the non-steady-state dynamics

The non-steady-state proton couple electron transport features tunable characteristic timescales. The rate constant $k_p$ and $k_n$, which are analogous to first-order reaction rates, are determined by the reaction mechanisms and the inherent properties of molecular ET. The nonlinearity of the voltage response in ET significantly alters the characteristic time ($\tau_{1/2} = \frac{1}{k_p + k_n}$), which is contingent upon the intrinsic transport mode and bias conditions. Meanwhile, variation in environmental water concentration directly affect the characteristic time. These shifts in the non-steady-state dynamics yield devices with intrinsically different dynamics, making them suitable for diverse input frequency scales. As depicted in figure 7a, an increase in atmospheric water concentration shifts the optimal input frequency peak for waveform recognition tasks to higher values, aligning the system's characteristic time with the input intervals of reservoir computing.

This characteristic behavior is akin to band-pass filtering properties in the classical signal processing system, showcasing the dynamic nature of this non-steady-state charge transfer system in the context of signal processing, as shown in figure 7b. It presents an interesting linkage between the dynamics of the system and signal frequencies in information processing, warranting further research to uncover more fundamental laws of chemical kinetics in information processing.

## Conclusion

In this study, we have explored the intricate dynamics of proton-coupled non-steady-state charge transport in the context of reservoir computing (RC), demonstrating its effectiveness in executing complex neural network tasks with a minimal set of dynamic molecular devices. Through careful adjustment of dynamic characteristics and input frequencies, our system displayed exceptional proficiency in waveform recognition and digit recognition tasks. In time series prediction, the system adeptly reconstructed attractors in the Hénon map and Lorenz weather system, showcasing its versatility.

Our investigation into how varying input frequencies influence the dynamic system as a reservoir has been particularly revealing. Crucially, we discovered that higher input frequencies lead to a diversification of reservoir states and reduced history dependence, highlighting the dynamic nature of the system. The interconnections and self-coupling of virtual nodes, key to fading memory effects, emerged as vital components in the effective functioning of RC. The characteristic time of the dynamic system, modulated by the types of chemical reactions and the molecules involved in electron transport, further influences the reservoir properties. Notably, water concentration (or general species B) in the environment serves as a pivotal factor for tuning the system, presenting a flexible model for in-materia computing.

Overall, our work leverages the unique attributes of non-steady-state chemical reaction dynamics to pioneer a reservoir computing model that seamlessly integrates dynamic kinetic and neural network performance. The ability to adjust characteristic time through the chemical atmosphere opens new avenues for controlling these systems. Additionally, the exploration of rich dynamic characteristics, including chaotic properties, offers exciting prospects for further research. Our findings aim to inspire innovative experimental approaches for constructing complex, functional neural network devices, harnessing the full potential of non-steady-state dynamics in information processing.

## ASSOCIATED CONTENT

Supporting Information
The Supporting Information is available free of charge via the Internet at http://pubs.acs.org.
Detailed methods for steady-state electron transport and proton transfer dynamics.

## AUTHOR INFORMATION

### Corresponding Author

Xi Yu – Tianjin Key Laboratory of Molecular Optoelectronic Science, School of Science, Tianjin University & Collaborative Innovation Center of Chemical Science and Engineering, Tianjin 300072, China;
orcid.org/0000-0001-5750-7003;
Email: xi.yu@tju.edu.cn.

### Author

Zhe-Yang Li – Tianjin Key Laboratory of Molecular Optoelectronic Science, School of Science, Tianjin University & Collaborative Innovation Center of Chemical Science and Engineering, Tianjin 300072, China;
Email: zheyangli@tju.edu.cn.

### Author Contributions

Z.L. developed the theoretical model and performed the reservoir computing. X. Y conceive the theoretical model and led the study. All the authors participated in manuscript writing and commenting on the manuscript. All authors have given approval to the final version of the manuscript.

Note
The authors declare no competing interest.

## ACKNOWLEDGMENT

This work is supported by the National Natural Science Foundation of China (21773169, 21973069), the PEIYANG Young Scholars Program of Tianjin University (2018XRX-0007), and Undergraduate Teaching Quality and Teaching Reform Project of Tianjin University (B201005607) and the Education Reform Project for Applied Chemistry Major (C20E06) in the Department of Chemistry of Tianjin University.